\documentclass[preprint]{aastex}

\usepackage{color}
\usepackage{natbib}
\usepackage{graphics}
\usepackage{graphicx}
\usepackage{amsmath}
\usepackage{url}

\begin{document}

\title{The Optimal Gravitational Lens Telescope}

\author{J. Surdej\altaffilmark{1}, C. Delacroix\altaffilmark{2}, P. Coleman\altaffilmark{3}, M. Dominik\altaffilmark{4}, S. Habraken\altaffilmark{2}, C. Hanot\altaffilmark{1}, H. Le Coroller\altaffilmark{5}, D. Mawet\altaffilmark{6}, H. Quintana\altaffilmark{7}, T. Sadibekova\altaffilmark{1}, D. Sluse\altaffilmark{8} }

\affil{$^1$ Li\`ege University, Department of Astrophysics, Geophysics and Oceanography (AGO), AEOS Group, All\'ee du 6 Ao\^ut 17, 4000 Li\`ege, Belgium}
\affil{$^2$ Li\`ege University, Department of Physics (DEPHY), Hololab Group, All\'ee du 6 Ao\^ut 17, 4000 Li\`ege, Belgium }
\affil{$^3$ University of Hawaii, Institute for Astronomy, 2680 Woodlawn Drive Honolulu, Hawaii 96822, USA }
\affil{$^4$ SUPA, School of Physics \& Astronomy, University of St Andrews, North Haugh, St Andrews, KY16 9SS, United Kingdom}
\affil{$^5$ Observatoire de Haute Provence, F-04870 Saint Michel l\textquoteright Observatoire, France }
\affil{$^6$ Jet Propulsion Laboratory, California Institute of Technology, 4800 Oak Grove Drive, CA 91109 Pasadena, USA }
\affil{$^7$ Departmento de Astronomia y Astrofisica, Pontificia Universidad Cat\'olica de Chile, Casilla 306, CL 22 Santiago, Chile}
\affil{$^8$ Astronomisches Rechen-Institut am Zentrum f\"ur Astronomie der Universit\"at Heidelberg, M\"onchhofstrasse 12-14, 69120 Heidelberg, Germany}

\email{surdej@astro.ulg.ac.be}

\shorttitle{The Optimal Gravitational Lens Telescope}
\shortauthors{Surdej et al.}

\begin{abstract}
Given an observed gravitational lens mirage produced by a foreground deflector (cf. galaxy, quasar, cluster, \dots), it is possible via numerical lens inversion to retrieve the real source image, taking full advantage of the magnifying power of the cosmic lens. This has been achieved in the past for several remarkable gravitational lens systems. Instead, we propose here to invert an observed multiply imaged source directly at the telescope using an ad-hoc optical instrument which is described in the present paper. Compared to the previous method, this should allow one to detect fainter source features as well as to use such an optimal gravitational lens telescope to explore even fainter objects located behind and near the lens. Laboratory and numerical experiments illustrate this new approach.

\end{abstract}


\maketitle

\section{Introduction}

 \citet{Zwicky37a,Zwicky37b} first proposed to use foreground galaxies as natural telescopes to observe otherwise too distant and faint background objects. The idea was either to take advantage of the gravitational lens amplification of the multiple unresolved images in order to obtain higher S/N observations of the background object(s) or to directly re-image, with a significantly improved angular resolution, the real extended source from the observed gravitational lens mirage. 
Numerical lens inversions have been successfully applied in the past to several cases among which the famous radio Einstein ring MG 1131+0456 \citep{Kochanek89}, the triply imaged giant arc in the galaxy cluster Cl 0024+1654 \citep{Wallington95}, the radio Einstein ring MG 1654+134 \citep{Wallington94,Wallington96}, the optical Einstein ring 0047-2808 \citep{Dye05},  the multiply imaged double source B1608+656 \citep{Suyu06}, the quadruply imaged quasar RXS J1131-1231 \citep{Claeskens06} or still, more recently, the large sample of lensed galaxies observed by the SLACS collaboration \citep{Bolton08}.  In the present paper, we show how to make use of an ad-hoc optical device at the telescope in order to directly invert an observed gravitational lens mirage.

First of all, we lay down in section 2 the basic principles of gravitational lensing inversion based upon straightforward ray tracing diagrams. We then show in section 3 how to use a point mass lens optical simulator to invert a gravitational lens mirage produced by a point mass deflector. After simulating in the laboratory an Einstein ring, a doubly or a quadruply imaged source, we show in section 4 how to invert the latter ones in order to retrieve the original source image. In section 5, we have made use of numerical simulations to confirm the expected properties of this type of hardware inversion. In section 6, we present a possible design for the optimal gravitational lens telescope and discuss the future prospects of gravitational lensing inversion using more sophisticated optical devices. Some general conclusions form the last section.

\section{Basic principles of gravitational lensing inversion}

\begin{figure*}[!ht]
\begin{center}
\includegraphics[width=15cm]{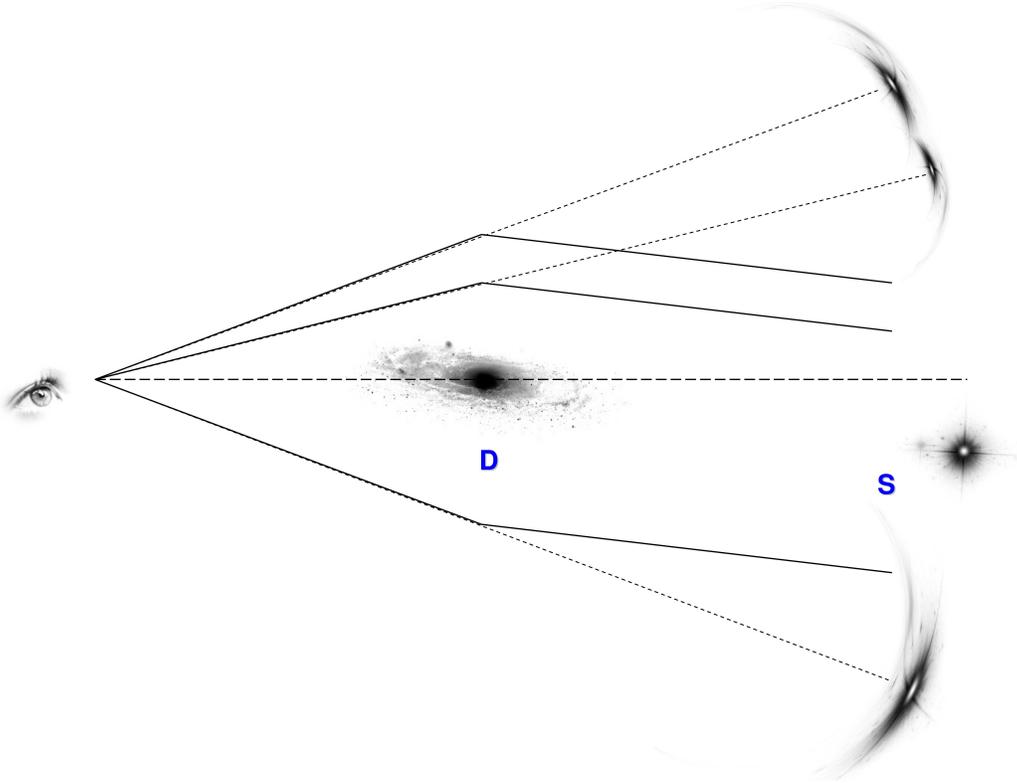}
\caption[]{Formation of a 3-lensed image mirage by gravitational lensing.}
\label{f1}
\end{center}
\end{figure*}

We have illustrated in Figure \ref{f1}  the possible deflection of light rays coming from a very distant source (S) by a foreground cosmic deflector (D, a galaxy here). We have assumed that in the present case, the observer located at left sees three lensed images of the distant source (S). 

\begin{figure*}[!ht]
\begin{center}
\includegraphics[width=15cm]{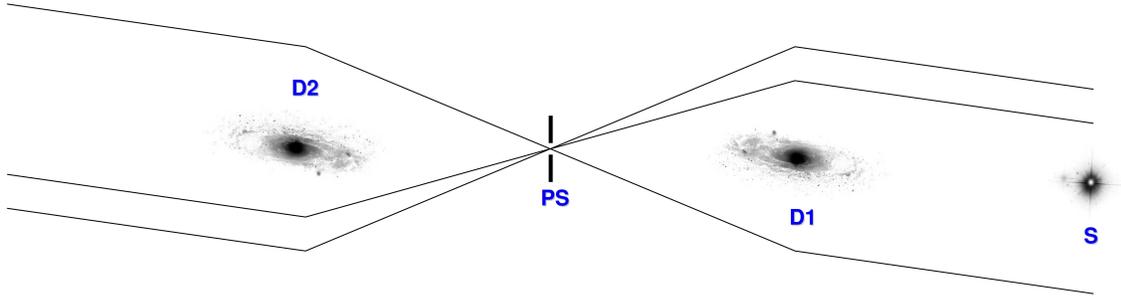}
\caption[]{This figure results from the superposition of the ray tracing diagram shown in Figure 1 and a copy of it, rotated by $180^\circ$. In this way, we clearly see that the light rays passing through the pinhole screen (PS) get similarly deflected, but in an opposite direction, to those coming from the distant source (S). The three beams of light rays passing the second, inverted, deflecting galaxy (D2) thus continue their travel as three parallel beams of light rays. }
\label{f2}
\end{center}
\end{figure*}

If one would set at the observer location three little mirrors, each parallel to the incoming wavefronts associated with each of the three beams of parallel light rays, the latter would travel back towards the source along their original trajectories in accordance with the principle of inverse travel of light, assuming of course the case of a static Universe.
Instead, let us suppose that at the location of the observer we just set a pinhole screen (PS) in order to only let the three incoming beams of parallel light rays continue their travels towards the left. The size of the pinhole is typically that of the telescope aperture used to directly image the cosmic mirages. Furthermore, let us place at a symmetric location with respect to the pinhole a galaxy (D2) similar to the original deflector (D1), but turned around by $180^\circ$. We then obtain the ray tracing diagram depicted in Figure \ref{f2}.

\begin{figure*}[!ht]
\begin{center}
\includegraphics[width=15cm]{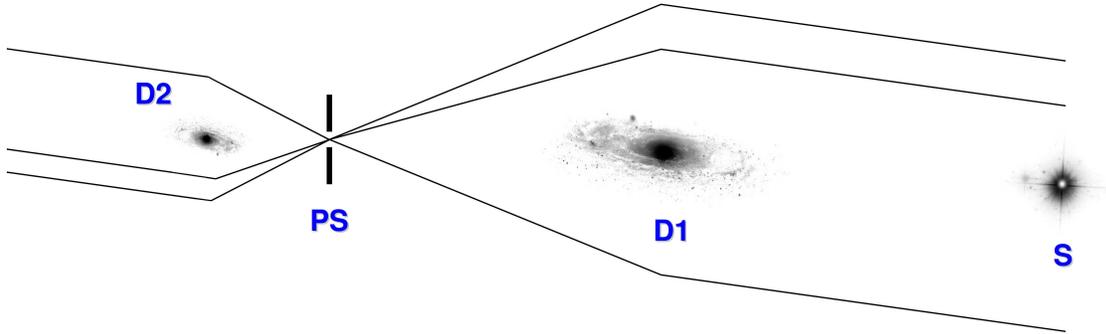}
\caption[]{Same as in Figure \ref{f2} but for the case of a deflector (D2) at left being located at a distance three times closer to the pinhole screen (PS) and being also three times less massive than the original deflector (D1). }
\label{f3}
\end{center}
\end{figure*}

Let us now assume that we could set the deflector at left (D2) much closer to the pinhole (PS, see Figure \ref{f3}). The deflector at left is still supposed to be perfectly aligned with the deflector at right (D1) and the pinhole. Let us suppose that the D2 deflector is now 3 times closer to the pinhole compared with the distance to the D1 galaxy. In order to preserve the same deflection angles $\alpha$ for the outgoing light rays at left, one simply needs to decrease the mass of the D2 deflector also by a factor 3, since $r$ has been reduced itself by the same factor, keeping of course identical the relative mass distribution between D2 and D1. Indeed, for the case of a symmetric mass distribution, we have $\alpha=4G M(r) c^{-2} \,r^{-1}$ where $M(r)$ represents the mass of the galaxy located within the impact parameter $r$; $G$ and $c$ being the universal constant of gravitation and the velocity of light \citep{Refsdal94}. If the ratio $M(r)/r$ is kept constant, the deflection angle $\alpha$ remains unaltered.

It is interesting to note that the ray tracing diagram shown in Figure \ref{f3} is very reminiscent of the propagation of light rays through a classical refractor, the main objective (cf. D1) being located at right from the pinhole screen (PS) and the eyepiece (D2) at left. Let us however note a big difference: in the present case, it is as if the main objective were diluted since only three (and not an infinity of) beams of parallel light rays propagate through such a gravitational lens refractor. Furthermore, these beams reaching the pinhole screen (or the observer) are mutually incoherent.

\begin{figure*}[!ht]
\begin{center}
\includegraphics[width=15cm]{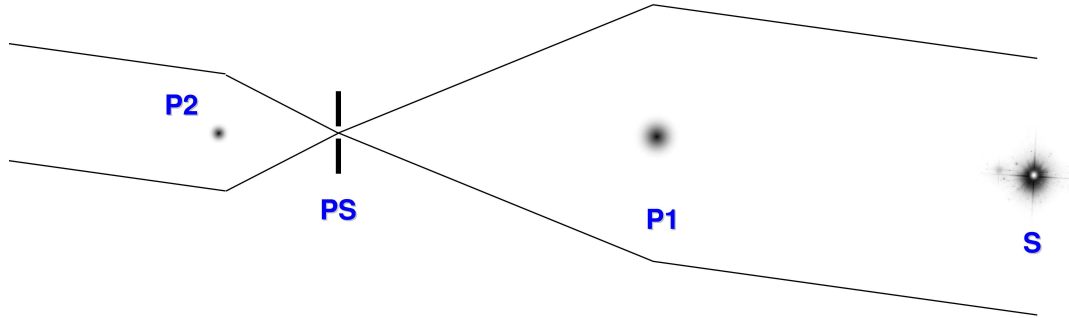}
\caption[]{A two lensed image mirage produced by a cosmic point mass lens (P1) at right. The two incoming beams of parallel light rays pass through the pinhole screen (PS) and get deflected by a point-mass lens object (P2) at left, at a distance 3 times closer from the pinhole and which mass is also 3 times smaller than that of the original cosmic point mass lens (P1). Alike for the case discussed in Figure \ref{f3}, the outgoing light rays are again parallel.}
\label{f4}
\end{center}
\end{figure*}

For convenience and simplicity, we shall consider in the remainder that the deflector is a point mass lens. In this case, the corresponding gravitational lens refractor is similar to that in Figure \ref{f3} with the exception that in general only two beams of parallel light rays go through the pinhole screen (see Figure \ref{f4}). Indeed, due to the singularity of this type of lens, the third Òcentral imageÓ is simply being suppressed \citep{Refsdal94}.

If we were now capable of setting the point mass lens object (P2) at left at a still much closer distance from the pinhole (PS), also decreasing accordingly its mass, the separation between the two outgoing parallel beams of light rays at left would get similarly smaller. Let us assume that this separation gets so small that it matches the size of an optical lens. 

Could we then replace the point mass lens object P2 by a laboratory lens simulator that would perfectly mimic the light deflection of the former? To our knowledge,  \citet{Liebes69} was the first to propose using an optical lens having the shape of the foot of a wine glass to simulate the light deflection by a point mass lens. The design and construction of such optical lenses corresponding to any type of axially symmetric mass distributions have been presented and discussed by  \citet{Refsdal94}. These authors have also described how to use such lenses to produce the various configurations of cosmic mirages observed in the Universe.

\section{The point-mass lens optical simulator and inversion of a gravitational lens mirage}

\begin{figure*}[!ht]
\begin{center}
\includegraphics[width=10cm]{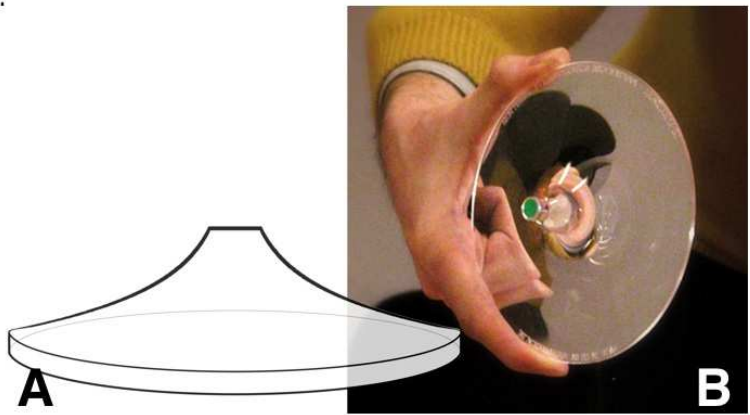}
\caption{Shape (a) and photograph (b) of an optical lens simulator corresponding to the case of a point mass lens. Such a lens is made of Plexiglas and its shape, similar to that of the foot of a wine glass, is essentially determined by the mass of the point mass deflector \citep{Refsdal94}. For the lens shown on the photograph, the corresponding mass is approximately 2/3 the mass of the Earth. }
\label{f5}
\end{center}
\end{figure*}

Figure \ref{f5} illustrates (a) the typical shape of the foot of a wine glass and (b) a corresponding simulator made of Plexiglas which has been used to simulate the light deflection of a point mass lens having approximately 2/3 the mass of the Earth when set at a distance of 1m from an observer eye.

\begin{figure*}[!ht]
\begin{center}
\includegraphics[width=15cm]{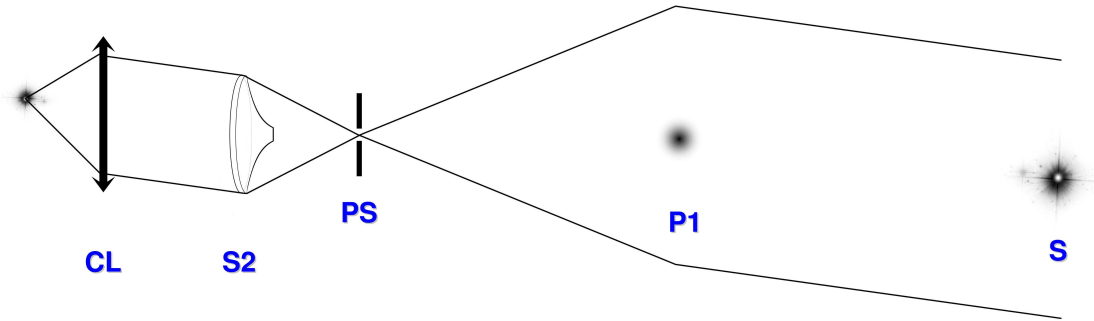}
\caption{Same as Figure \ref{f4} but the small size cosmic lens at left (P2) has been replaced by an optical point mass lens simulator (S2) corresponding to the same mass. Since the separation between the two outgoing parallel beams of light rays is now reduced to several tens of centimetres, or even smaller, it is easy to place at left a classical converging lens (CL) so that a perfect, lens inverted, image of the distant source at right (S) is formed in its focal plane. }
\label{f6}
\end{center}
\end{figure*}

We may now conveniently replace in Figure \ref{f4} the small size cosmic point mass lens (P2) at left by an optical point mass gravitational lens simulator (S2) characterized by the same mass (see Figure \ref{f6}).

Still further at left, we set a converging lens (CL) so that a perfect image of the distant source (S) is formed in its focal plane. The real gain in proceeding so is to observe the recombined image of the distant source with a significant increase in angular resolution, corresponding to the highest magnification(s) provided by the point mass cosmic lens (P1), at right.
A simple laboratory experiment illustrating these principles is proposed in the next section.

\section{The laboratory gravitational lens experiment for the case of a point mass lens}

\begin{figure*}[!ht]
\begin{center}
\includegraphics[width=15cm]{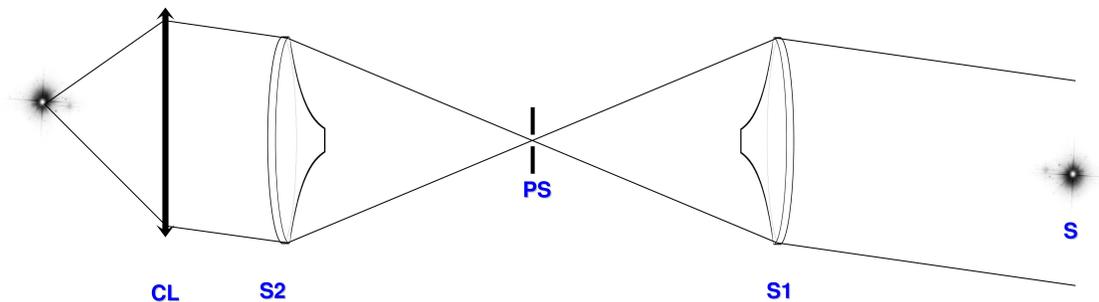}
\caption{The first point mass gravitational lens simulator (S1) at right produces a doubly imaged source as seen from the pinhole (PS) while the second lens simulator (S2) inverts the mirage into two parallel beams of light rays which are then focused at left by a classical converging lens (CL).}
\label{f7}
\end{center}
\end{figure*}

\begin{figure*}[!ht]
\begin{center}
\includegraphics[width=15cm]{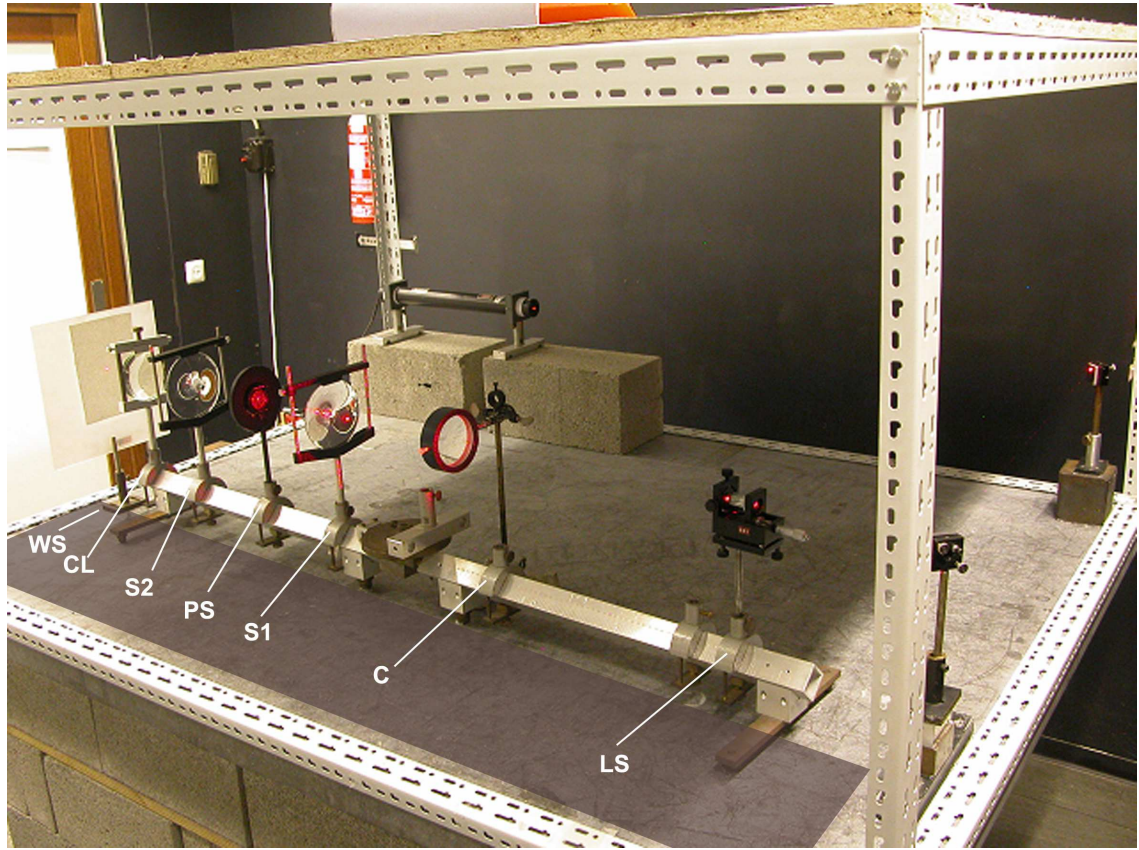}
\caption{Optical bench in the laboratory showing from right to left the laser point source (LS) obtained with a spatial filter (microscope objective combined with a pinhole screen), followed by a lens (C) that collimates the light rays into a parallel beam which enters the first point mass gravitational lens simulator (S1). Some of the deflected rays then encounter the pinhole screen (PS) and enter the second optical lens simulator (S2). The outgoing light rays are finally focused by means of a converging lens (CL) on the white screen (WS) set at the extreme left. }
\label{f8}
\end{center}
\end{figure*}

\begin{figure*}[!ht]
\begin{center}
\includegraphics[width=15cm]{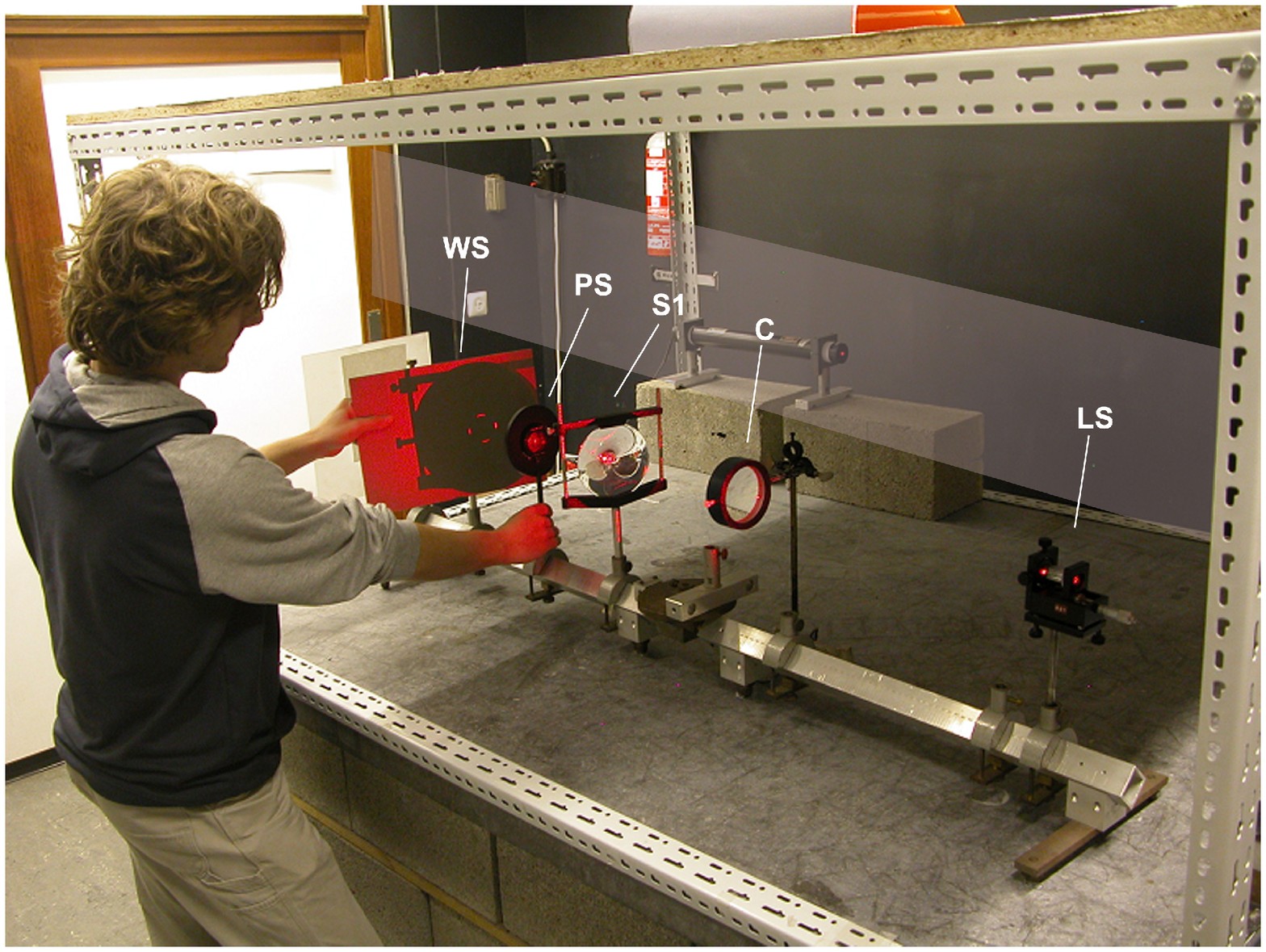}
\caption{Another view of the laboratory optical bench (see also Figure 8). In this case, the first point mass gravitational lens simulator (S1) has been somewhat tilted with respect to the axis of the optical bench. By placing a white screen (WS) just behind the pinhole screen (PS), one sees the formation of a quadruply imaged source, in accordance with the predictions made for the case of a point mass lens in presence of an external shear \citep{Refsdal94}. }
\label{f9}
\end{center}
\end{figure*}

For purely didactical purposes, the Faculty of Sciences from the Li\`ege University has proceeded to the manufacturing of a large series of optical gravitational lens simulators alike the one shown in Figure \ref{f5} (b) (see \url{http://www2.ulg.ac.be/sciences/lentille/dp-lentilleweb.pdf}). In order to illustrate the principles exposed in the previous section, we have thus decided to use two such optical lenses in the laboratory: one (S1) to produce a mirage and the second one (S2) to invert the resulting lensed images. The corresponding ray tracing diagram is shown in Figure \ref{f7} while the laboratory experiment is shown in Figures \ref{f8} and \ref{f9}.

\begin{figure*}[!ht]
\begin{center}
\includegraphics[width=15cm]{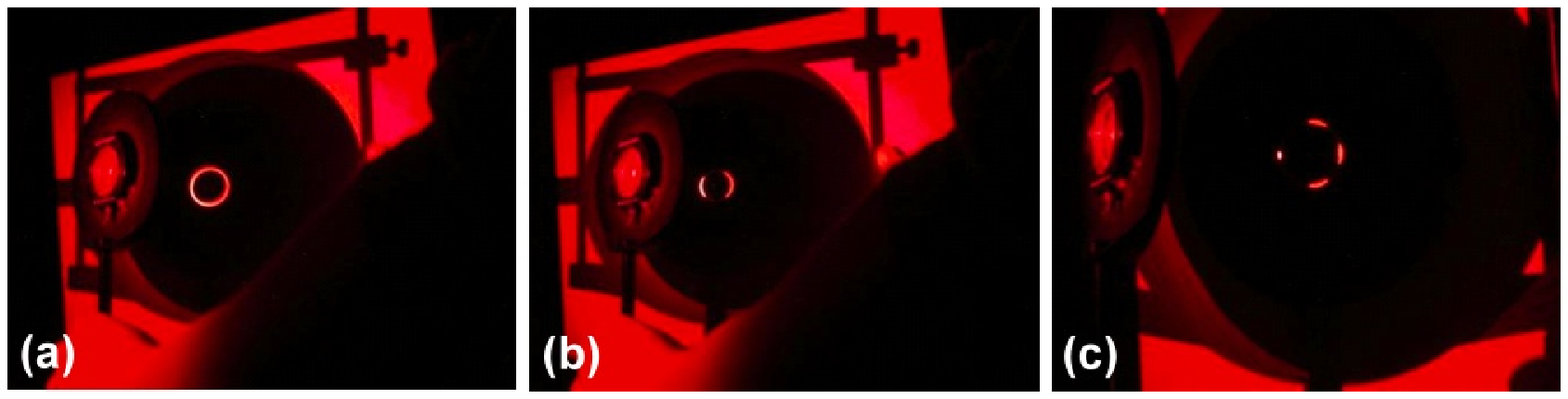}
\caption{In case of perfect alignment between the source (LS), the pinhole (PS) and the optical lens (S1), set perpendicularly with respect to the axis of the optical bench, there results the formation of an Einstein ring (a) as seen on the white screen (WS) set between the pinhole (PS) and the S2 lens (cf. Figure 9). If we slightly translate the S1 lens, along a direction transverse to the axis of the optical bench, the Einstein ring breaks in two lensed images (b). If instead, we slightly tilt the S1 lens with respect to the axis of the optical bench, the Einstein ring breaks into four lensed images (c) \citep{Refsdal94}.  }
\label{f10}
\end{center}
\end{figure*}

\begin{figure*}[!ht]
\begin{center}
\includegraphics[width=15cm]{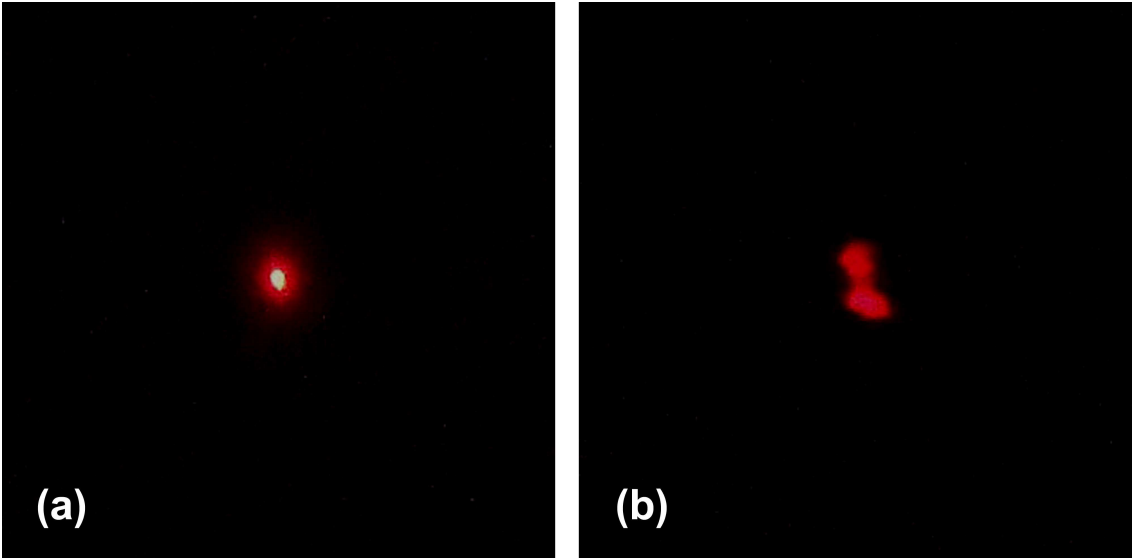}
\caption{Inverted lensed images of the laser source on the white screen (WS) placed at the extreme left in Figure 8, for the case of a single (a) point-like source image S and that of a double one (b). The double source was obtained by means of a beam splitter. The diameter of the spot(s) seen in (a) and (b) is approximately 2mm. In these two cases, the S1 and S2 lenses were set perpendicularly to the optical bench axis, slightly translated in opposite transverse directions and symmetrically placed with respect to the pinhole (PS). }
\label{f11}
\end{center}
\end{figure*}

Various experiments have been carried out in the laboratory, by tilting (or not) and/or translating (or not) with respect to each other the two optical point mass gravitational lens simulators S1 and S2, also for the case of a double point-like laser source image S. We have illustrated in Figure \ref{f10} some of the resulting lensed images seen on the intermediate white screen (WS) placed between the pinhole screen (PS) and the S2 lens (cf. Figure \ref{f9}). Finally, keeping always the S1 and S2 lenses parallel to each other and symmetrically placed with respect to the pinhole, we have observed for various configurations the images formed on the white screen (WS) placed at the extreme left position on the optical bench as shown in Figure \ref{f8}. In all cases, we either observe (see Figure \ref{f11}) the formation of a single point-like image or that of a double one, in case the original source (LS) consisted of a double point-like laser source. We have thus succeeded in inverting the multiple lensed images produced by the S1 point mass lens simulator into the original, single or double, laser source image.

\section{Numerical simulations}

\begin{figure*}[!ht]
\begin{center}
\includegraphics[width=15cm]{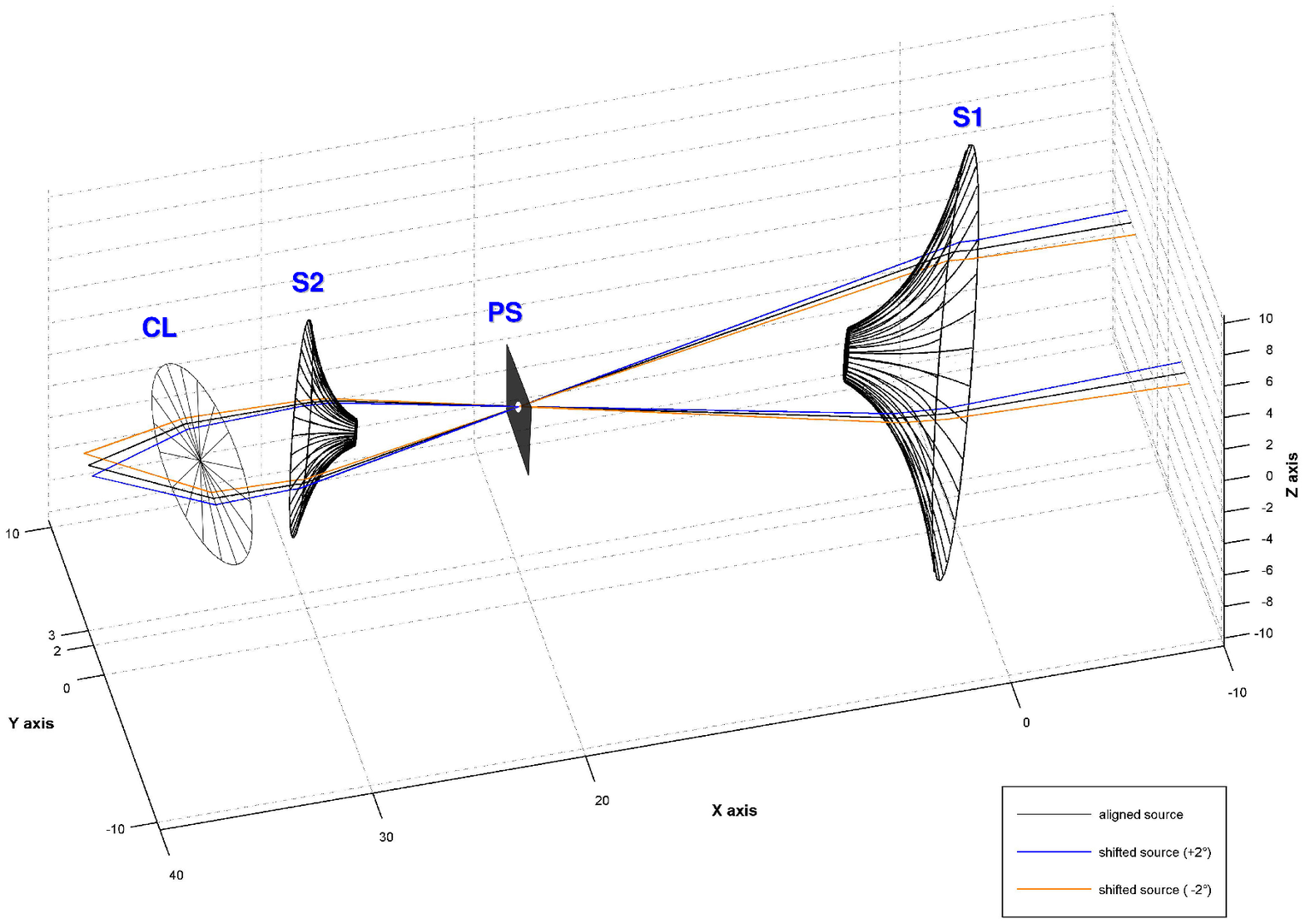}
\caption{3-D ray tracing simulations performed with the Matlab software for the case of two similar optical point mass lens simulators (S1 and S2) which centers have been properly aligned with the pinhole (PS). The two simulators have been kept parallel to each other but slightly tilted ($10^\circ$) with respect to the axis of the optical bench.}
\label{f12}
\end{center}
\end{figure*}

\begin{figure*}[!ht]
\begin{center}
\includegraphics[width=15cm]{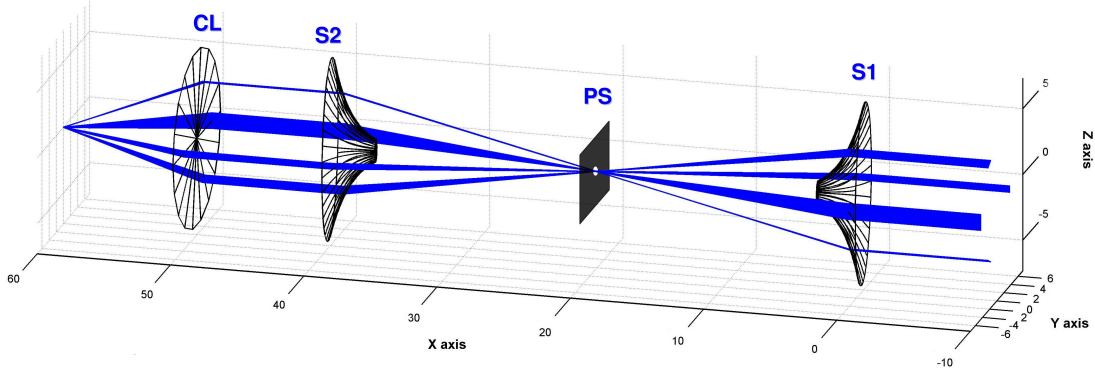}
\caption{Another example of 3-D ray tracing simulations performed with the Matlab software for the case of the two optical point mass lens simulators (S1 and S2) which centers have been properly aligned with the pinhole (PS). The two simulators have been kept parallel to each other but slightly tilted ($20^\circ$) and translated along a transverse direction with respect to the axis of the optical bench. Note that one of the lenses (S2) is half as massive as the other one (S1) and placed at half the distance with respect to the pinhole in order to provide an overall optimal gravitational lens telescope.  }
\label{f13}
\end{center}
\end{figure*}

Matlab software applications have been developed to simulate the propagation of light rays through two optical point mass lens simulators alike those used in the laboratory. The ray tracings have been performed considering the refractive properties of the optical simulators (shapes, refractive index n = 1.49, etc.). Numerical 2D and 3D simulations have been carried out for various tilts and translations of the S1 and S2 simulators, also for different sizes of these lenses and their positions with respect to the pinhole. Typical examples of such simulations are illustrated in Figures \ref{f12} and \ref{f13}. Such simulations have been carried out for various source positions as well as for the case of multiple point-like sources. In all cases, the inversions by the optical gravitational lens simulator S2 of the mirages produced by the simulator S1 have properly restored in the focal plane of a converging lens CL the original source images.

\section{The optimal gravitational lens telescope: design and future prospects}

\begin{figure*}[!ht]
\begin{center}
\includegraphics[width=10cm]{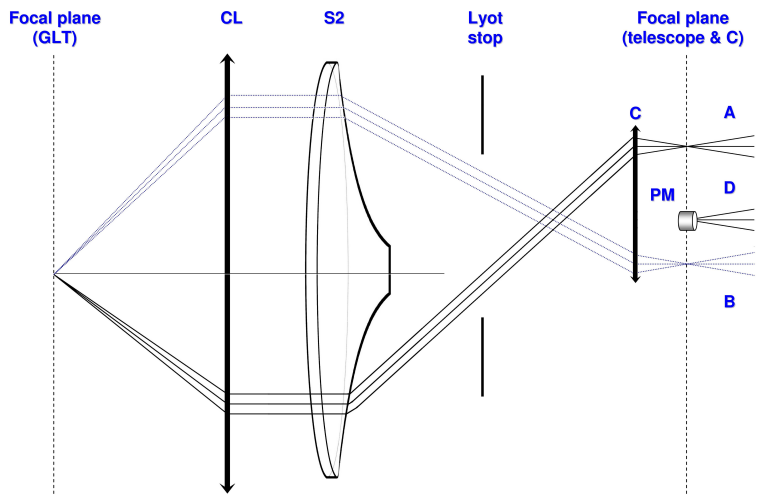}
\caption{A simple unfolded model of the optimal gravitational lens telescope. It is here assumed that the astronomical telescope pointed towards a gravitational lens system produces in its focal plane at right two lensed images (A \& B) of a distant source as well as the image of the deflector (D). The latter has been centred on the Lyot or phase mask (PM) of the coronagraphic device shown in this figure. The lens C collimates the light, forming also an exit pupil where a Lyot stop is placed. The optical point mass lens simulator S2 then inverts the observed gravitational lens mirage into two parallel beams of light rays which are then focused at left by a classical converging lens (CL). A single image of the original source, originally observed as a multiply imaged object, has thus been restored in the focal plane of CL. }
\label{f14}
\end{center}
\end{figure*}

We propose to directly invert an observed gravitational lens mirage using an ad-hoc optical instrument placed at the focus of a large telescope as follows. The idea is first to extinguish as much as possible the direct light from the foreground deflector which might be a galaxy, a star, a quasar, etc. To do so, we centre the light of the deflector on the mask of a coronagraphic device. The mask could either be a classical Lyot one if the foreground deflector is resolved or a phase mask alike an annular groove phase mask in case the deflector is point-like \citep{Mawet05b}, see Figure \ref{f14}. Just behind, a lens (C) collimates the light rays from the observed mirage (a doubly imaged point-like source for the case shown in Figure \ref{f14}) and forms an exit pupil of the telescope aperture where a Lyot stop is placed. The optical point mass lens simulator (S2), or another ad-hoc optical device adapted to the specific mass distribution of the lens, then inverts the mirage and produces two (or more) parallel beams of light rays. Finally a classical converging lens (CL) produces in its focal plane a single image of the multiply imaged distant source. This is a possible design for the optimal gravitational lens telescope which ideally restores in its focal plane a single real and faithful image from the direct imaging of a multiply imaged source. 

As can be seen in Figure \ref{f14}, only very small regions of the lens simulator (S2 in the present case) are actually involved in the lens inversion process. One could therefore think of possibly replacing this lens simulator by a computer generated holographic lens or even a deformable mirror such that the entropy of the recombined image in the focal plane of the converging lens gets minimized. 
Conversely, one could use the lens model inferred from the numerical lens inversion of a known gravitational lens system to design the corresponding optical lens inversion simulator. Direct imaging of the gravitational lens mirage with such an optical device would then permit the direct observation of the original source image as well as to possibly detect still much fainter objects located behind and near the foreground deflector. Compared to the numerical inversion method, the sensitivity should be higher since the light from the source is now concentrated on a smaller number of pixels compared to the spread of the multiple images over a larger number of pixels. This is specially true whenever the noise in the faint lensed images is dominated by the CCD read-out-noise. Moreover, the coronagraphic device removes the light from the deflector that would otherwise be spread over the detector. As far as the final angular resolution is concerned, we should select a pixel size and/or adapt the focal length of the CL lens (see Figure 14) in such a way that the most magnified image of the source is not undersampled. 

For the particular case of nearby stars assuming that their mass and distance are precisely known, one could directly design optical lens simulators alike those shown in Figure 5 and search for very faint and distant background objects located behind and near these stellar cosmic lenses. Note that some of the cosmological lenses may turn out to have a complex structure. For these, the corresponding optical lens inversion simulators ought to be more sophisticated. 

Many other applications might be envisaged.

\section{Conclusions}

We have shown in the present paper how Zwicky's proposition \citep{Zwicky37a,Zwicky37b} to use foreground deflectors as giant cosmic lenses could be directly achieved at the telescope, using a phase mask coronagraph equipped with an ad-hoc optical lens simulator in order to invert in real time an observed gravitational lens mirage into its real source image. The resulting optimal gravitational lens telescope is thus simply composed of the cosmic gravitational lens producing the observed cosmic mirage, the observer's telescope and a coronagraphic lens inversion instrument placed at its focus. 
Based upon the gravitational lens modelling of known resolved mirages, it should be straightforward to design the corresponding optical lens simulators associated with the mass distribution of the corresponding deflectors in order to directly invert the former ones at the telescope. The use of such ad-hoc instruments should allow one to directly re-image the real source at much fainter levels as well as to detect even fainter background objects located behind and near the foreground deflector(s). 
We have the aim to design and test in the near future such an instrument for the case of the well resolved quintuply imaged quasar SDSS J1004 + 4112 at z = 1.734 produced by a foreground cluster at z=0.68 \citep{Inada03,Liesenborgs09}. We also intend to test the use of deformable mirrors as well as computer generated holograms in order to properly restore in the focal plane of the optimal gravitational lens telescope the real image source(s) of observed gravitational lens mirages. 

\acknowledgements
The authors wish to dedicate the present work to the memory of Prof. Sjur Refsdal (1935-2009) who has been a pioneer in the field of gravitational lensing theory. Sjur Refsdal has also been a very keen friend and collaborator. The authors from the Li\`ege University acknowledge support from the Communaut\'e fran\c{c}aise de Belgique - Actions de recherche concert\'ees - Acad\'emie universitaire Wallonie-Europe and thank the Faculty of Sciences (Li\`ege University), in particular Prof. J.-M. Bouquegneau, for making possible a mass production of gravitational lens simulators. D.S. acknowledges the support of a Humboldt Fellowship and H.Q. partial support from the Belgian F.R.S.-FNRS during his sabbatical stay in the Li\`ege University and the FONDAP Centro de
Astrofisica.

\bibliographystyle{apj}
\bibliography{aeos_bib}

\end{document}